\documentclass{elsart}
\usepackage{latexsym}
\usepackage{epsfig}
\usepackage{natbib}

\def\url#1{#1} 

\begin{document}

\begin{frontmatter}

\title{The conditions for coherent scattering of starlight in the forward 
direction, and it's consequences
}

\author{Fr\'ed\'eric \snm Zagury}
\address{
\cty 02210 Saint R\'emy Blanzy, \cny France
\thanksref{email} }

   \thanks[email]{E-mail: fzagury@wanadoo.fr}

\accepted{May 2002}
\communicated{F. Melchiorri}

 \begin{abstract}

     I have been too conservative in `A review of the properties of the scattered starlight which 
contaminates the spectrum of reddened stars' (NewA, 2002, 7/4, 191) concerning the 
conditions implied by an additional component of scattered light in the 
spectrum of reddened stars. 
I will review and simplify these conditions.
Implications for the column density of the scatterers will then be
discussed.
Estimates for coherent Rayleigh scattering by hydrogen are 
in excellent agreement with the observations.
     
\end{abstract} 
\end{frontmatter}
 \section{Introduction} \label{intro}
The conditions I have set in `A review of the properties of the scattered starlight which 
contaminates the spectrum of reddened stars' \citep{uv8} for scattered 
starlight to represent an important proportion of the spectrum of a 
reddened star are too conservative.
In this paper I will review and simplify these conditions (section~\ref{cond}).
I will then look further into how the angular distance from the reddened 
star effects the scatterers contribution.
In section~\ref{interfer}, the conditions found in section~\ref{cond} are used to see the 
implications for the column density and the nature of the scatterers.
\section{Conditions for coherent scattering} \label{cond}
Two conditions out of the three set in \citet{uv8} cannot be avoided.
Firstly, the scatterers have to be identical and spherically symetric so that the amplitudes
of the scattered waves will be equal and add positively.
Secondly, the phase lag between the scattered waves has to be small, which 
implies that the scatterers are very close to the reddened star 
direction:
the order of magnitude of the difference between the path followed by the scattered 
waves and the star-observer distance is given by the UV wavelengths ($\sim 1000\rm\AA$).

Interferometry with a large source of light, a candle for instance or, 
at a larger scale, a star, requires the product of the dimension of the 
source by  the distance  between the receptors to be small compared to the 
product of the wavelength by the distance between the source and 
the receptors \citep{berkeley}. 
This condition, applied to the scattering of starlight by interstellar 
particles on the line of sight of a star, is a third and major constraint on 
the conditions for coherent scattering given in \citet{uv8}.
It is more restrictive than the 
condition which was set on the optical paths followed by the scattered 
waves.
In the particular case of the light scattered in the complete forward 
direction, this third condition can be proved not to be necessary.

Consider the star as the set of the many point sources  which 
altogether give the overall light of the star.
For one such source, $s$, the particles (supposed to be identical and 
spherically symetric) very close to the source-observer 
axis scatter the source's light with negligible phase lag.
The intensity of the scattered starlight received by the observer is proportional 
to $N_{s}^{2}$, the square of the column density of scatterers on the 
point-source line of sight: $I_{s}= i_{s}*N_{s}^{2}$, with $i_{s}$ the 
intensity the observer would receive from one particle alone.

Since $N_{s}=N$ should not differ from one point source to the other, 
the intensity of the scattered light received by the 
observer is proportional to $N^{2}$.

If condition~1 from \citet{uv8} is released, coherent scattering 
certainly occurs from the particles within an angle $\theta$ (viewed from 
the observer) from the star which satisfies equation~2 of \citet{uv8}:
\begin{equation}
    \theta  \ll 
    5\,10^{-8}"\left(\frac{\lambda}{2000\,\rm\AA}\right)^{0.5}
     \left(\frac{100\,pc}{l_{0}}\right)^{0.5}
     \left(\frac{d_{0}}{D}\right)^{0.5}
    \label{eq:c1}
\end{equation}¥
$l_{0}$ is the observer-cloud distance, $d_{0}$ the star-cloud 
distance, $D$ the star-observer distance ($D=d_{0}+l_{0}$).

A typical order of magnitude for the 
angular distance from the star within which scattered starlight is 
coherent is  $10^{-8}"$,  a larger value than the $10^{-12}"$ found in \citet{uv8}.
For a cloud at $100\,\rm pc$ this represents $\sim 10^{14}\,\rm cm^{2}$ 
(a radius of $\sim 100\,\rm km$).
If $\beta$ is the proportion of scatterers relative to the number of 
hydrogen atoms, for a typical $N_{H}=10^{21}\,\rm H/cm^{2}$ column 
density, coherent scattered light is, at the first order, $\beta 
10^{35}$ times what it would be if scattering was incoherent.  
\section{Further contribution of scatterers at larger distances from 
the star-observer axis} \label{interfer}
Particles at larger distances from the 
star-observer axis will give a scattered wave with a larger phase-lag, and can be thought to give a negative contribution to the scattered 
starlight. 
In \citet{uv4}, I thought this could explain the $2200\,\rm\AA$ bump.

Such contribution will not annhilate the positive ones, but will 
change the wavelength dependence of the scattered starlight.
This is seen by integrating all contributions, in a calculus similar to 
that given by \citet{vandehulst}, section~4.3.

The $1/\lambda^{4}$ dependence observationally found for the scattered 
light \citep{uv3, uv6} indicates that
the contribution of destructive waves is probably negligible.

The most obvious reason is that, when moving away from the line of sight 
of the star, the spatial distibution of the 
scatterers becomes an important factor to consider in the addition of the amplitudes of 
the scattered waves.
Coherent scattering at too large distances from the star direction will occur only in specific, ordered configurations, which cannot 
be the case in interstellar space.
Since we compare differences of astronomical distances 
to wavelengths of less than one micron, deviations from the 
first Fresnel zones introduce random phase lags which annihilate the 
contribution of these waves \citep{bohren}.

Another reason might come from Section 2 if interferences of the scattered waves implies that relation 1 of Zagury (2002c) be respected.
Since this relation is e.g. much more restrictive than 
relation~\ref{eq:c1} of section~\ref{cond}, interferences will 
concern particles which, anyway, do not introduce phase lags between 
the scattered waves.
\section{The column density of the scatterers} \label{coldens}
In section~6.1 of \citet{uv8}, I have distinguished two possible kinds of 
scatterers.
The scatterers can be atoms or molecules from the 
gas, the most natural and likely candidate being hydrogen.
Small dust particles (the VSG introduced by 
\citet{sellgren}) will also scatter as $1/\lambda^{4}$.
The scattering cross section of each of these two possible candidates 
have different expressions, which I will use to examine how likely each 
carrier can be responsible for the coherent scattering.

The maximum ratio of scattered light to direct starlight corrected for 
extinction and measured on earth, is of order $15\%$.
It is reached for $\lambda\sim 1500\,\rm\AA$ \citep{uv2}.

From equation~7 of \citet{uv4} the order of magnitude of the ratio  (scattered 
light)/(direct starlight corrected for extinction) can be estimated to be:
\begin{equation}
    \frac{\sigma(NS)^{2}}{4\pi l_{0}^{2}}
    \left(\frac{D}{d_{0}}\right)^{2}
    \sim
    \sigma N^{2}\theta^{4}l_{0}^{2}\left(\frac{D}{d_{0}}\right)^{2}¥
    \label{eq:fluxdif}
\end{equation}¥
with $S$ the surface of the cloud around the star-observer axis 
within which coherent scattering occurs, 
$\sigma=\sigma_{0}(2000\,{\rm \AA}/\lambda)^{4}$ ($\sigma$ is given in 
cm$^2$ in the following) the optical depth of the scatterers for 
Rayleigh scattering.

A ratio of $15\%$ between the scattered light and the direct 
starlight corrected for extinction is obtained for:
\begin{equation}
    N=  4\, 10^{-10} \left(\frac{1''}{\theta} \right)^{2}
     \left(\frac{d_{0}}{D} \right)
      \left(\frac{100\,{\rm pc}}{l_{0}} \right)
       \left( \frac{\lambda}{2000\,{\rm \AA}}\right)^{2}¥
        \sigma_{0}^{-0.5}\,\mathrm{cm}^{-2}
    \label{eq:n}
\end{equation}
With $\lambda=1500\,\rm\AA$ and condition~\ref{eq:c1} we find a simple condition on $N$:
\begin{equation}
    N\gg 10^{7} \sigma_0^{-0.5}\,\mathrm{cm}^{-2}
    \label{eq:condn}
\end{equation}
$\sigma_{0}$ is given for hydrogen and for the VSG in \citet{uv8}.

For Rayleigh scattering by hydrogen: $\sigma_{0}=2.3\,10^{-28}$.

For small spherical grains: 
$\sigma_{0}=3.2\,10^{-8}(a/100\,\rm nm)^{6}$ 
(\citet{vandehulst}, section~6.4).
$a$ is the size of the particles.

Thus, depending on whether the scatterers are hydrogen or small grains, they must 
satisfy:
\begin{eqnarray}
    N_{H}& \gg & 6\,10^{20}\,\mathrm{cm}^{-2}
    \label{eq:nh}  \\
    N_{grains}& \gg & 10^{7}
    \left(\frac{100\,\mathrm{nm}}{a}\right)^3
    \,\mathrm{cm}^{-2} 
    \label{eq:ng}
\end{eqnarray}
Condition~\ref{eq:nh} is in remarkable agreement with the column densities found 
for hydrogen in interstellar clouds.

From the \citet{bohlin} relation $A_V/N_H \sim 5\,10^{-22}\,\rm mag/cm^2$, 
an hydrogen column density $N_H\sim 6\,10^{20}\,\rm cm^{-2}$ 
corresponds to $A_V\sim 0.3$, and $E(B-V)\sim 0.1$ (assuming $R_{V}\sim 3$).
$E(B-V)\sim 0.1$ is also the average observed value above which scattered starlight starts to represent an 
appreciable part of the spectrum of reddened stars \citep{uv5, uv7}.
The hypothesis, suggested by T.~Lehner (Observatoire de Meudon), that hydrogen 
is the carrier responsible for the contamination of starlight 
by scattered light, leads to an agreement between theory and observatons.

Condition~\ref{eq:ng} is flexible, if identical and symetric VSG do 
exist, they can also be the agents of the coherent scattering.
\section{Conclusion} \label{conc}
I have reviewed conditions for coherent scattering of 
starlight by interstellar particles.
This review leads to substantial corrections of sections~5 and 6.2 of 
\citet{uv8}.

The conditions specified in this paper concern the nature of the scatterers 
and the trajectories followed by the scattered waves. 
The length of these optical paths must differ by less than a wavelength,
leading to one mathematic condition, equation~\ref{eq:c1} of section~\ref{cond}.

The order of magnitude of the angular angle from the reddened star within which starlight is 
coherently scattered is $\sim 10^{-8}$''.
For a cloud at $100\,$pc this represents a surface of radius $\sim 100\,$km.

We have observational indications that only the scatterers very 
close to the star direction participate to the coherent scattering, 
since, if this was not the case, the $\lambda$ dependence of the scattered 
light will have a different power law than $1/\lambda^{4}$.
This rules out the possibility of interferences due to waves 
scattered at larger distances from the star-observer axis,
conforming to the idea that the random distribution of the 
scatterers eliminates coherent scattering when 
moving away from the star direction \citep{bohren}. 

The scatterers must be small particles, atoms or molecules, or small 
grains.
Conditions on the column density of the scatterers have also been set 
for each of these types of particles.

Atomic hydrogen can seriously be considered to be the carrier of the scattered light which 
contaminates the spectrum of reddened stars.
The column densities of hydrogen necessary to provide the amount of scattered 
starlight are typical of the interstellar medium. 
Furthermore, the lower limit found for the hydrogen column density 
agrees with the value of $E(B-V)$, given by observation \citep{uv5, uv7}, 
above which the scattered light component becomes noticeable in the spectrum of reddened stars.

Note that, reciprocally, section~\ref{coldens} indicates that coherent Rayleigh scattering by hydrogen must give an appreciable 
contribution to the spectrum of reddened stars.
{}
\end{document}